\documentclass[12pt]{article}
\usepackage{graphicx}
\usepackage{color}
\usepackage{cite}
\usepackage{here}

\def \beq{\begin{equation}}
\def \eeq{\end{equation}}
\def\eqref#1{(\ref{#1})}
\def\bea{\begin{eqnarray}}
\def\eea{\end{eqnarray}}
\def\jpsi{J\kern-0.1em/\kern-0.1em\psi\kern0.03em}

\def\URLtilde{\lower0.2em\hbox{$\tilde{\phantom{a}}$}}
\def\mycomm#1{\hfill\break\strut\kern-3em{\color{red}\tt ====> #1
\color{black}}\hfill\break}

\newcount\timecount
\newcount\hours \newcount\minutes  \newcount\temp \newcount\pmhours
\hours = \time
\divide\hours by 60
\temp = \hours
\multiply\temp by 60
\minutes = \time
\advance\minutes by -\temp
\def\hour{\the\hours}
\def\minute{\ifnum\minutes<10 0\the\minutes
\else\the\minutes\fi}
\def\clock{
\ifnum\hours=0 12:\minute\ AM
\else\ifnum\hours<12 \hour:\minute\ AM
\else\ifnum\hours=12 12:\minute\ PM
\else\ifnum\hours>12
\pmhours=\hours
\advance\pmhours by -12
\the\pmhours:\minute\ PM
\fi
\fi
\fi
\fi
}

\def\monthname{\relax\ifcase\month 0/\or January\or February\or
March\or April\or May\or June\or July\or August\or September\or
October\or November\or December\else\number\month/\fi}

\def\bold#1{\setbox0=\hbox{$#1$}     \kern-.025em\copy0\kern-\wd0
\kern.05em\copy0\kern-\wd0
\kern-.025em\raise.0433em\box0 }

\begin{document}
\setcounter{footnote}{1}
\rightline{EFI 21-5}
\rightline{arXiv:2107.04915}
\vskip1.0cm

\centerline{\large \bf Configuration mixing in strange tetraquarks $Z_{cs}$}
\bigskip

\centerline{Marek Karliner$^a$\footnote{{\tt marek@tauex.tau.ac.il}}
 and Jonathan L. Rosner$^b$\footnote{{\tt rosner@hep.uchicago.edu}}}
\medskip

\centerline{$^a$ {\it School of Physics and Astronomy}}
\centerline{\it Raymond and Beverly Sackler Faculty of Exact Sciences}
\centerline{\it Tel Aviv University, Tel Aviv 69978, Israel}
\medskip

\centerline{$^b$ {\it Enrico Fermi Institute and Department of Physics}}
\centerline{\it University of Chicago, 5620 S. Ellis Avenue, Chicago, IL
60637, USA}
\bigskip
\strut

\begin{quote}
\begin{center}
ABSTRACT
\end{center}
The BESIII Collaboration has observed a candidate for a $c \bar c s \bar u$
tetraquark $Z_{cs}$ at $(3982.5^{+1.8}_{-2.6} \pm 2.1)$ MeV and width
$(12.8^{+5.3}_{-4.4} \pm 3.0)$ MeV, while the LHCb Collaboration has observed
a $Z_{cs}$ candidate in the $\jpsi K^-$ channel with mass of
$(4003 \pm 6 ^{+4}_{-14})$ MeV and width $(131 \pm 15 \pm 26)$ MeV.  In this
note we examine the possibility 
that these two states are distinct eigenstates of a mixing
process similar to that which gives rise to two axial-vector mesons
labeled by the Particle Data Group \cite{PDG} $K_1(1270)$ and $K_1(1400)$.
The main point is that on top of a $\bar c c$ pair,
the $Z_{cs}$ states have the same light quark content as the $K_1$-s.
In the compact tetraquark picture this implies several additional 
states, analogous to members of the $K_1$ nonet. These states 
have not yet been observed, nor are they required in the molecular 
approach. Thus experimental discovery or exclusion of these extra states 
will be a critical test for competing models of exotic mesons with hidden
charm.
\end{quote}
\smallskip

\leftline{PACS codes: }
\bigskip


\section{INTRODUCTION \label{sec:intro}}

The discovery of mesons with minimal quark configuration $c \bar c q \bar q'$,
where $q$ and $q'$ are light quarks $u,d,s$, was pioneered by the observation
\cite{Choi:2003ue} of the $X(3872)$.  The quark configuration of this state
includes $c \bar c u \bar u$, $c \bar c$, and some lesser admixtures, while
in terms of mesons it is mainly $D^0 \bar{D}^{*0} +$ charge conjugate (c.c.)
The spin, parity, and charge-conjugation eigenvalue of this state is 
$J^{PC} = 1^{++}$ \cite{JPC}.

Over the course of the nearly twenty subsequent years, partners of this state
have emerged. Within the compact tetraquark model this implies
the existence of a flavor $SU(3)$ nonet whose neutral
members have $J^{PC} = 1^{++}$, as well as candidates for membership in a
$1^{+-}$ nonet \cite{Maiani:2004vq,Maiani:2021tri,Giron:2021sla,Giron:2020qpb,
Lebed:2016yvr}.\footnote{Strictly speaking, $1^{++},1^{+-}$
denotes the {\it neutral} member of a nonet in which the light quark-antiquark
pair is in a $^3\!P_1,{^1\!P_1}$ state. 
We shall label these basis states $(A,B)$.}
Candidates for the strange members of these nonets have been identified.
The BESIII Collaboration has observed a candidate for a $c \bar c s \bar u$
tetraquark $Z_{cs}$ at $(3982.5^{+1.8}_{-2.6} \pm 2.1)$ MeV and width
$(12.8^{+5.3}_{-4.4} \pm 3.0)$ MeV \cite{Ablikim:2020hsk}, while the LHCb
Collaboration has observed a $Z_{cs}$ candidate in the $\jpsi K^-$ channel
with mass of $(4003 \pm 6 ^{+4}_{-14})$ MeV and width $(131 \pm 15 \pm 26)$ MeV
\cite{Aaij:2103ivw}.

A number of interpretations of the $Z_{cs}$ candidates have been put forth
\cite{Maiani:2004vq,Yang:2020nrt,Ortega:2021enc,Meng:2021rdg}, reviewed
briefly in \cite{Karliner:2021}.  Here we propose that 
within the compact tetraquark model these two states are likely to be
distinct eigenstates of a mixing process similar to that which gives rise to
two axial-vector strange mesons labeled by the Particle Group \cite{PDG}
$K_1(1270)$ and $K_1(1400)$.  We discuss the system of these two kaons in
Sec.\ \ref{sec:mixk}, apply similar methods to the $Z_{cs}$ states in Sec.\
\ref{sec:mixz}, and conclude in Sec.\ \ref{sec:conc}.

\section{MIXING OF STRANGE KAONS \label{sec:mixk}}

We label quark-antiquark mesons with the notation $^{2S+1}L_J$, where $S=0,~1$
is the total quark spin, $L = S,~P,~D, \ldots$ stands for orbital angular
momentum $0,~1,~2, \ldots$, and $J = 0,~1,~2, \ldots$ denotes total angular
momentum.  The lowest-level quark-antiquark mesons consist of

\beq \label{eqn:sw}
\hbox{$S$-waves:}\ J^{PC} = 0^{-+}=~^1\!S_0,~~1^{--}=~^3\!S_1;~
\eeq
\beq \label{eqn:pw}
\hbox{$P$-waves:}\ J^{PC} = 0^{++}=~^3\!P_0,~~1^{++}=~^3\!P_1,~~
                         1^{+-}=~^1\!P_1,~~2^{++}=~^3\!P_2~~.
\eeq
Here $P=\pm 1$ stands for parity eigenvalue, while $C=\pm 1$ denotes eigenvalue
under charge conjugation, which is only a good quantum number for a $q \bar q'$
pair if $q' = q$.  More generally we shall use the basis labels $(A,B)$ for
    $(^3\!P_1,{^1\!P_1})$, respectively.

The assignment of some of the observed light mesons to the above quantum numbers
is given in Table 15.2 of Ref.\ \cite{PDGQM}.  The two lightest nonets
of $J^P=1^+$ mesons are also listed in Table~\ref{tab:P1}.

\begin{table}[H] 
\caption{Lowest-lying nonets of $J^P = 1^+$ mesons}
\label{tab:P1}
\begin{center}
\begin{tabular}{c c c c c c} \hline \hline
$^{2S+1}L_J$ & $J^{PC}$ & $I=1$ & $I=1/2$ & $I=0$ & $I=0$ \\ \hline
 $^3\!P_1$ & $1^{++}$ & $a_1(1260)$ & $K_{1A}^\dag$ & $f_1(1420)$ &
   $f_1(1285)$ \\
 $^1\!P_1$ & $1^{+-}$ & $b_1(1235)$ & $K_{1B}^\dag$ & $h_1(1415)$ &
   $h_1(1170)$ \\ \hline \hline
\end{tabular}
\end{center}
 \leftline{\hskip 1in $^\dag$ $\!K_1(1270)$ and $K_1(1400)$ are mixtures of
$K_{1A,B}$ (see text).}
\end{table}

Nonstrange states, in the limit
of isospin conservation, possess a definite $G$-parity, defined as
\beq
G \equiv C \exp(i \pi I_2)
\eeq
which reduces to $G = C (-1)^I$ for the neutral member of an isospin multiplet
with isospin~$I$.  However, because the strange quark is heavier than the
up and down quarks, the corresponding symmetry based 
on the $U$-spin and $V$-spin 
\cite{Lipkin:1964zza} subgroups of $SU(3)_F$  does not exist  
(though an analogue of $G$-parity can be formally defined
in the $SU(3)_F$ symmetry limit \cite{Dothan:1963cyc}).

Evidence for two resonant states with $J^P = 1^+$ (axial vector kaons) in the
$K\pi\pi$ spectrum was first noted in Ref.\ \cite{Goldhaber:1967}.  The mass
eigenstates arise from mixing between $K_{1A}={^3\!P_1}$ and $K_{1B}={^1\!P_1}$,
for instance as a result of sharing $K^*(892)\pi$ and $K \rho$ decay
modes \cite{CandR,mixref}.  They may be written as \cite{CandR} (Ref.\
 \cite{Tayduganov:2011} uses a different phase convention)
\bea
|K_1(1270) \rangle&=&|K_{1A}\rangle \cos\phi - |K_{1B} \rangle \sin\phi~,\\
|K_1(1400) \rangle&=&|K_{1A}\rangle \sin\phi + |K_{1B} \rangle \cos\phi~.
\eea

Both $K_1(1270)$ and $K_1(1400)$ can couple to $K^*(892)\pi$ and $K \rho$, 
in $S$ and $D$ waves.  Ref.\ \cite{Tayduganov:2011} quotes parameters of Ref.\
\cite{ACCMOR} for these decays as shown in Table \ref{tab:ACCMOR}, while
current PDG values including those obtained from nondiffractive axial kaon
production are summarized in Table \ref{tab:PDGdec}.
%
\begin{table} 
\caption{Masses and decay properties of axial vector kaons as measured by the
ACCMOR Collaboration \cite{ACCMOR} in diffractive production and quoted in
Ref. \cite{Tayduganov:2011}.  Small $D$-wave contribution to $K_1(1270) \to
K \rho$ neglected.}
\label{tab:ACCMOR}
\begin{center}
\begin{tabular}{c c c c c c} \hline \hline
 $K_1$ & $M$, MeV & $\Gamma_{K_1}$, MeV & ${\cal B}(K^*\pi)_S$ &
  ${\cal B}(K^*\pi)_D$ & ${\cal B}(K \rho)_S$ \\ \hline
 $K_1(1270)$ & $1270 \pm \phantom{2}7$ & $\phantom{1}90 \pm \phantom{3}8$ 
& $0.13 \pm 0.03$ & $0.07 \pm 0.006$
 & $0.39 \pm 0.04$ \\
$K_1(1400)$ & $1410 \pm 25$ & $165\pm35$ & $0.87 \pm 0.05$ & $0.03 \pm 0.005$
 & $0.05 \pm 0.04$ \\ \hline \hline
\end{tabular}
\end{center}
\end{table}
%
\begin{table} 
 \caption{Masses and decay properties of axial vector kaons as summarized in
Ref.\ \cite{PDG}, including nondiffractive production.  Small $D$-wave
contribution to $K_1(1270) \to K \rho$ neglected.}
\label{tab:PDGdec}
\begin{center}
\begin{tabular}{c c c c c c} \hline \hline
$K_1$ &  $M$, MeV & $\Gamma_{K_1}$, MeV & ${\cal B}(K^*\pi)_S$ &
  ${\cal B}(K^*\pi)_D$ & ${\cal B}(K \rho)_S$ \\ \hline
 $K_1(1270)$ & $1253 \pm 7$ & $\phantom{1}90 \pm 20$ & $0.08 \pm 0.025^\dag$ &
 $0.08 \pm 0.025^\dag$ & $0.42 \pm 0.06$ \\
$K_1(1400)$ & $1403 \pm 7$ & $174\pm 13$ & $0.90 \pm 0.06\kern0.9em\strut$ 
& $0.04 \pm 0.01\kern0.9em\strut$ & $0.03 \pm 0.03$ \\ \hline \hline
\end{tabular}
\end{center}
\leftline{$^\dag$Neglecting error in ${\cal B}(K^*\pi)_D/{\cal B}(K^*\pi)_S
= (1.0 \pm 0.7)\%$}
\end{table}
The analysis of Ref.\ \cite{CandR} favored the range $10^0 < \phi <
35^0$.  For that range, the mixing between mass eigenstates and basis
states $|K_{1A,B} \rangle$ is such that
$K_1(1270)$ couples mainly to $K \rho$ in S wave, with suppressed S wave
coupling to $K^* \pi$ despite greater phase space, while $K_1(1400)$ couples
mainly to $K^* \pi$ in S wave.  This pattern is qualitatively similar to that
in an updated version \cite{Tayduganov:2011,mixref}.  
Mixing of states with similar quantum numbers such that different eigenstates
have different decay modes is familiar from many other examples in hadron
spectroscopy \cite{Koniuk:1980, Capstick:1993,Karliner:2007}.

\section{MIXING OF $Z_{cs}$ STATES \label{sec:mixz}}

Within the compact tetraquark model it is natural to expect that
the $Z_{cs}$ candidates at 3982.5 and 4003 MeV are the
analogues of $K_1(1270)$ and $K_1(1400)$ with an extra $c \bar c$ pair.
The light-quark degrees of freedom are linear combinations of a $^3P_1$
and a $^1P_1$ $q \bar q'$ pair.  
This scheme is close to a superposition of Solutions 1 and 2 of Ref.\
\cite{Maiani:2021tri}, in which $Z_{cs}(3982.5)$ and $Z_{cs}(4003)$ are distinct
states of $^{2S+1}P_J$: $^{3,1}P_1$ for Solution 1
and $^{1,3}P_1$ for Solution 2.  In our solution, it is the the unmixed
strange $^3P_1$ state which should have a mass $M_A$ which is the average of
the nonstrange $X(3872)$ and the $c \bar c s \bar s$ state called $X(4140)$:
\beq \label{eqn:mplus}
 M_A = (4146.8 + 3871.65)/2~{\rm MeV} = 4009~{\rm MeV}~,
\eeq
where we have used the latest PDG masses
and the identification by Lebed and Polosa \cite{Lebed:2016yvr} of the
$X(4140)$ as a $1^{++}$ $c \bar c s \bar s$ state.  In addition to
the $X(3872)$, assumed isosinglet, there should exist a nearby nonstrange
isotriplet, whose charged members await detection.  Strictly speaking, if
we regard the $X(3872)$ as mainly $c \bar c u \bar u$, thanks to its proximity
to the $D^0 \bar{D}^{*0}$ + c.c. threshold, there should also exist a state
which is mainly $c \bar c d \bar d$, presumably not far from the $D^+
\bar{D}^{*-}$ threshold, so the estimate (\ref{eqn:mplus}) is uncertain by at
least several MeV.  One should stress that while the existence of these
additional states is a must in the compact tetraquark picture, in the molecular
approach they are not necessary and in some cases downright unlikely.

In particular, $X(3872)$ mass is extremely close, within 0.1 MeV of
the \,$\bar D^0 D^{*0}$\, threshold \cite{LHCb:2020xds}. One pion exchange
likely plays a significant role in $\bar D^0 D^{*0}$ binding, converting a
\,$\bar D^0 D^{*0}$\, system into a \,$\bar D^{*0} D^0$, which of course has the
same mass. On the other hand, in the putative charged partner of $X(3872)$
one pion exchange would have to convert a \,$\bar D^0 D^{*+}$, 
with threshold at $3875.1$ MeV to a \,$D^{*0} D^+$, with threshold at 
3876.5 MeV, a shift of 1.4 MeV, much too large in comparison with the 
dynamics apparently responsible for $X(3872)$ binding.

In view of these conflicting expectations from the compact tetraquark and
molecular pictures,
experimental search for the extra states is of very high priority.
It can discriminate between these competing models of exotic mesons with
hidden charm.  A discovery would provide strong support for the compact
tetraquark picture, while definite experimental exclusion would 
be a very strong argument in favor of molecular models.

In what follows we shall refer to nonets by the light-quark spin-parity
$^{(2S+1)}L_J$ of their unmixed members, so we will speak of the $^3\!P_1$
and $^1\!P_1$ nonets.  Regarding decay modes of the $^3\!P_1$ nonet, 
those of the $X(3872)$ and $X(4140)$ are not affected by $K_1$-like mixing, 
because the light quark content of these states is $\bar u u$ and $\bar s s$,
respectively.  On the other hand, we propose that as result of the mixing
of the $^1\!P_1$ and $^3\!P_1$ multiplets $Z_{cs}(3982.5)$ is the eigenstate of
$c \bar c s \bar u$ mixing which couples to $D_s^- D^{*0} + D_s^{*-} D^0$,
while $Z_{cs}(4003)$ is the eigenstate which couples to $\jpsi K^-$.

Information about the $^1\!P_1$ nonet is more sparse.  The only firm candidate
is the nonstrange isotriplet $Z_c(3900)$, whose latest mass listed by the
PDG is $(3887.1 \pm 2.6)$ MeV.  The $c \bar c s \bar s$ candidate with
$J^{PC} = 1^{+-}$ is unknown at present.  That prevents us from estimating the
singly strange member of the nonet with mass $M_B$ from the data. 
If all members of the $^1\!P_1$ nonet are $M(Z_c(3900))-M(X(3872)) \sim 15-20$
MeV above those of the $^3\!P_1$ nonet, we predict a $C=-1$ $c \bar c s \bar s$
candidate around 4160 MeV.  (See also \cite{Giron:2021sla,Maiani:2021tri,%
Giron:2020qpb,Lebed:2016yvr}.)  It should be able to decay to $D_s^{*-} D_s^+
+ D_s^- D_s^{*+}$, whose threshold is 4080 MeV.

\section{CONCLUSIONS \label{sec:conc}}

We have proposed that the two strange tetraquarks recently observed by
the BESIII and LHCb Collaborations, $Z_{cs}(3982.5)$ and $Z_{cs}(4003)$,
are mass eigenstates in which the singly strange light quark degrees of freedom
with $^{2S+1}\!L_J = {^{3,1}\!P_1}$ undergo mixing such that the lighter state
couples predominantly to $D_s^- D^{*0} + D_s^{*-} D^0$, while the heavier
one couples mainly to the mode in which it was observed, $\jpsi K^+$.  This
mixing is similar to that giving rise to the axial vector kaons $K_1(1270)$
and $K_1(1400)$, which has been known for more than fifty years.  The scheme
envisions both states as having $J^P = 1^+$. It predicts a $^1\!P_1$ state
of negative $C$ around 4160 MeV, decaying to $D_s^{*-} D_s^+ + D_s^- D_s^{*+}$.
The compact tetraquark scheme discussed here involves complete nonets; in a
molecular picture key nonet members (such as the charged partners of the
$X(3872)$) may be absent.

\section*{ACKNOWLEDGMENTS}

The research of M.K. was supported in part by NSFC-ISF grant No.\ 3423/19.
We thank S. Godfrey for helpful correspondence and R. Lebed for calling
our attention to Ref.\ \cite{Giron:2021sla}, which also draws the analogy
with axial vector kaon mixing but considers $Z_{cs}(3985)$ and $Z_{cs}(4000)$
the same and takes the second state to be $Z_{cs}(4220)$.

\end{document}